\documentclass{revtex4-1}

\usepackage{graphicx}%
\usepackage{dcolumn}%
\usepackage{bm}%
\usepackage{epsfig}
\usepackage{latexsym}
\usepackage{epstopdf}

\begin{document}


\preprint{LPT-ORSAY 13-81}

\title{Crystal phases of soft spheres systems in a slab geometry} 



\author{Moritz Antlanger}
\email[]{moritz.antlanger@tuwien.ac.at}
\affiliation{Institut f\"ur Theoretische Physik and Center for Computational Materials Science (CMS), Technische Universit\"at Wien, Wiedner Hauptstra{\ss}e 8-10, A-1040 Wien, Austria} 
\affiliation{Laboratoire de Physique Th\'eorique (UMR 8627), Universit\'e de Paris-Sud and CNRS, B\^atiment 210, F-91405 Orsay Cedex, France} 
\author{G\"unther Doppelbauer}
\affiliation{Institut f\"ur Theoretische Physik and Center for Computational Materials Science (CMS), Technische Universit\"at Wien, Wiedner Hauptstra{\ss}e 8-10, A-1040 Wien, Austria}
\author{Martial Mazars}
\email[]{martial.mazars@th.u-psud.fr}
\affiliation{Laboratoire de Physique Th\'eorique (UMR 8627), Universit\'e de Paris-Sud and CNRS, B\^atiment 210, F-91405 Orsay Cedex, France}
\author{Gerhard Kahl}
\affiliation{Institut f\"ur Theoretische Physik and Center for Computational Materials Science (CMS), Technische Universit\"at Wien, Wiedner Hauptstra{\ss}e 8-10, A-1040 Wien, Austria}


\date{\today}

\begin{abstract}
We have identified the ground state configurations of soft particles (interacting via inverse power potentials) confined between two hard, impenetrable walls. To this end we have used a highly reliable optimization scheme at {\it vanishing} temperature while varying the wall separation over a representative range. Apart from the expected layered triangular and square structures (which are compatible with the three dimensional bulk fcc lattice), we have identified a cascade of highly complex intermediate structures. Taking benefit of the general scaling properties of inverse power potentials, we could identify -- for a given softness value -- one single master curve which relates the energy to the wall separation, irrespective of the density of the system. Via extensive Monte Carlo simulations, we have performed closer investigations of these intermediate structures at {\it finite} temperature: we could provide evidence to which extent these particle arrangements remain stable over a relatively large temperature range.

\end{abstract}


\maketitle 

\section{Introduction}
The prediction of ordered equilibrium structures for a system of particles interacting via a specific (pair) interaction is a highly non-trivial and, in the general case, a so far unsolved problem\cite{Maddox:88}. The search for its comprehensive and thus satisfactory solution represents a formidable challenge which is not only of purely academic interest, but also of paramount relevance in numerous branches of condensed matter physics. Already the probably most simple example, namely the packing of hard spheres, provides an insight into the complexity of the problem: Kepler's suggestion of a hexagonal close-packed arrangement as the optimal arrangement of hard spheres \cite{Kepler:11} could only be proven rigorously a few years ago\cite{Hales:05}. On this occasion, one should also mention a comprehensive study on the ordered phases formed by soft particles (interacting via an inverse power potential), considering different degrees of softness\cite{Agrawal:95}.

The search for ordered equilibrium structures becomes even more intricate when these configurations have to be identified under the condition that the system is subject to geometric constraints due to an external confinement. Again, even the most simple problem of this type, namely hard spheres confined between two parallel, unstructured and non-interacting walls, still lacks a comprehensive solution, despite considerable efforts that have been dedicated to this problem in the past years\cite{Oguz:09,Oguz:12}. Since only packing arguments are responsible for the formation of the ordered equilibrium structures for a fixed wall separation $H$, the complexity of this particular problem can easily be illustrated: only for those (discrete) $H$-values that correspond to the thickness of $n_l$ parallel slices of the fcc bulk crystal is the solution of the problem obvious, leading to a stack of $n_l$ triangular or square layers (denoted in the following by $n_l \triangle$ and $n_l \Box$, respectively). For all other $H$-values, the particles arrange in such a way as to optimize the packing, leading to a cascade of highly intricate intermediate structures, such as prismatic, buckling or rhombic phases\cite{Ramiro:09}. 

Even though several studies have been dedicated to this particular problem in recent years, complete sequences of these intermediate structures over a representative range of $H$-values could be identified only recently ({\it cf.} Refs.\ [\onlinecite{Oguz:09},\onlinecite{Oguz:12}] and references therein). If, in addition, energy comes into play -- {\it e.g.} via the softness in the repulsion or via an attractive tail -- the situation is, of course, even more intricate; this complexity has been exemplified by the few systems that have been considered in literature, so far: Gaussian particles \cite{Constantin:08,Pansu:11,Kahn:09,Kahn:10}, particles interacting via exponentially decaying potentials \cite{Fortini:06}, Lennard-Jones systems \cite{Bock:05,Gribova:11}, and water models\cite{Zangi:04} are a few examples.

The properties of systems confined between two parallel plates are also relevant for many experimental systems such as colloids confined in thin wedges\cite{Pieranski:80,Pieranski:83,Neser:97,Ramiro:06,Ramiro:09,Schoen:book:07} and nanoparticles trapped in surfactant bilayers\cite{Constantin:08,Pansu:11}. The crossed arrangement of the two cylinders in a surface force apparatus\cite{Israelachvili:book:92} induces a typical slab geometry. With this setup, layering transitions can be studied as the distance between the cylinders is varied at constant pressure.

In this contribution we focus on particles that interact via inverse power potentials. The paramount relevance of such systems stems from their widespread use as repulsive reference potentials in dense liquids for the study of static, thermodynamic and dynamical properties\cite{Hoover:71, Hoover:72,Pedersen:08,Pedersen:10,Coslovich:08,Agrawal:95,Hansen:13}. When used in repulsive reference systems, inverse power potentials represent effective interactions and the exponent $n$ can, for instance, be estimated from the scaling properties of the diffusion coefficients\cite{Coslovich:08} or by applying the virial theorem\cite{Pedersen:08,Pedersen:10}. The values for the exponents computed in this manner are in general greater than ten.

For this contribution, we have investigated the crystal phases of a system of soft spheres confined between two parallel walls, separated by a distance $H$. The spherical particles interact via an inverse power potential for which we have considered different degrees of softness. We have identified the ordered ground state configurations at {\it vanishing} temperature with optimization tools based on ideas of evolutionary algorithms, varying the wall separation over several particle diameters. In complementary Monte Carlo simulations we have tested the mechanical stability of some of these particle arrangements at {\it finite} temperature. With increasing distance between the plates, we could identify a sequence of alternating, layered triangular and square structures, which we have denoted by $1\triangle$, $2\Box$, $2\triangle$, $3\Box$, etc.\ (see notation introduced above). In relatively narrow $H$-intervals, where the system creates a new layer or where a transformation from a $\triangle$- to a $\Box$-phase (or vice versa) takes place, a cascade of highly complex, intermediate structures could be identified, some of which remain stable at finite temperature.

This paper is organized as follows. In the subsequent section we present the model system that we have investigated and briefly summarize the methods used in this study: an optimization tool based on ideas of evolutionary algorithms and Monte Carlo simulations. Section \ref{sec:results} is dedicated to a detailed discussion of the results; particular emphasis is put on the structure and on the stability of four intermediate phases; all the intermediate ground state configurations that have been identified in our investigations are summarized in the Supplementary Material\cite{ref:supplementary} of this contribution. The paper closes with concluding remarks.

\section{Model and methods}
\label{sec:models_methods}

\subsection{Model} 
\label{subsec:model} 

In the present work we investigate systems of confined soft particles, interacting via a repulsive inverse power law (IPL) interaction

\begin{equation}
\label{eq:IPL}
\displaystyle V_n(r)=4\epsilon \left(\frac{\sigma}{r}\right)^n ;
\end{equation}
$r$ is the distance between particles, $\epsilon$ and $\sigma$ set the energy-and length-scales, respectively. In the limit $n\rightarrow\infty$, we recover the hard sphere potential, where $\sigma$ corresponds to the diameter of the impenetrable particles.

The softness of the interaction \cite{Agrawal:95} between particles can be defined as $1/n$. For large values of $n$, the interaction is short-ranged; in contrast, it is long-ranged in monolayer and slab systems for $n \leq 2$ and in bulk systems for $n \leq 3$. In the present work, we focus throughout on short-ranged interactions: for all computations reported in this work, $n \geq 12$.

\begin{figure}
\begin{center}
\includegraphics[width=7cm]{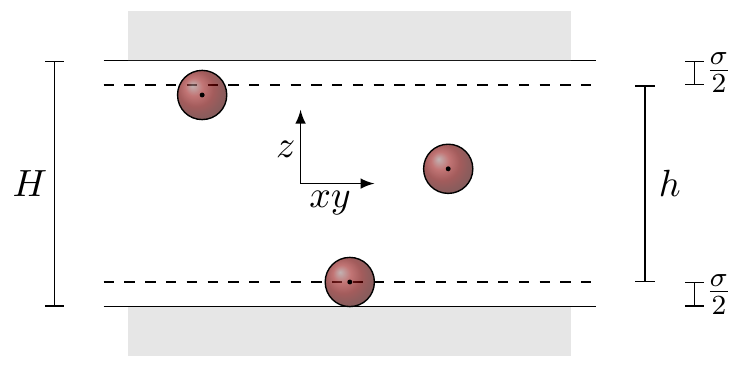}
\end{center}
\caption{\label{fig:sketch} Schematic representation of the geometry of the system investigated in this contribution. The centers of the particles (with diameter $\sigma$) are confined in a slab of effective width $h$. The impenetrable walls are separated by a distance $H=h+\sigma$. The position of the particles, ${\bf r}$, can be decomposed into ${\bf r} = {\bf s} + z \hat {\bf e}_z$, ${\bf s}$ being a two dimensional vector.}
\end{figure}

In this contribution we consider systems with slab geometry as shown schematically in Figure \ref{fig:sketch}: particles are confined between two parallel, impenetrable walls, which are separated by a distance $H = h + \sigma$. $h$ is thus the effective slab width, {\it i.e.}, the range that is actually accessible to the particle centers. The $z$-axis is chosen perpendicular to the plate and the origin of the coordinate system is fixed in the center of the slab; thus the confining walls are located at $z = \pm H/2$. Assuming a cell as a finite sub-volume of the slab (which is either the crystal unit cell in the optimization runs or the simulation box in simulations), its volume is given by $V_H=AH$ and the volume accessible to the particles is $V_h=Ah$, $A$ being the base area of the cell.  The case $h = 0$ corresponds to the purely two-dimensional monolayer, while the case $h\rightarrow\infty$ is equivalent to the three-dimensional bulk; in all other cases, we deal with a quasi-two-dimensional system. In all computations specified in Subsection \ref{subsec:methods} ({\it i.e.}, both in optimizations based on evolutionary algorithms (EA) and in Monte Carlo (MC) simulations) periodic boundary conditions are applied in directions parallel to the confining walls.

The number density in the slab, $\rho_h = N/Ah$, with $N$ being the number of particles per cell, is trivially related to the number density $\rho_H = N/AH = h\rho_h/(h+\sigma)$. In the following we set $\rho = \rho_h$, a quantity which will henceforward be given in units of $1/\sigma^3$. The average distance between two neighboring particles can be estimated via the Wigner-Seitz radius $r_0=(3/4\pi \rho)^{1/3}$. Therefore, a natural parameter to study properties of slab systems is the ratio $h/r_0$.

Since we consider in the present work only short-ranged interactions, it is convenient to introduce a cutoff radius $R_{\rm{cut}}$ when computing the energy of the system via lattice sums; its particular value, $R_{\rm{cut}} \simeq 10 \sigma$, is chosen such that the relative accuracy for the energy is better than $10^{-6}$; the use of Ewald summation techniques is thus not required \cite{Mazars:11,Mazars:10}.

In order to quantify the structures occurring in the quasi two-dimensional geometry of the model, the use of suitably adapted order parameters is required: we decided to modify the commonly used two-dimensional, $m$-fold rotational order parameters\cite{Steinhardt:83,Mickel:13,Voronoi:Book} via a factor that includes the vertical distance $z_{ij}$ between particles. Thus, for particle $i$ we have defined the local, modified order parameter as

\begin{equation}
\label{eq:local_boop}
\Psi_m(i) = \frac{1}{\tilde{W}_{\rm{b}}(i)} \sum_{j=1}^{N_{\rm{b}}(i)} 
\exp\left(-\frac{z_{ij}^2}{\left( \alpha r_0\right)^2}\right) 
\exp\left(\imath m \theta_{ij}\right) ,
\end{equation}
where $\theta_{ij}$ is the angle of the bond between particles $i$ and $j$ with respect to a reference direction (in our case the $x$-axis).  Further, $\tilde{W}_{\rm{b}}(i)$ is the normalization factor that corresponds to the total number of neighbors $N_{\rm{b}}(i)$ of particle $i$, modulated by a Gaussian weight associated with the vertical distance to the neighbor $j$; $\tilde{W}_{\rm{b}}(i)$ is thus defined as

\begin{equation}
\label{eq:norm_boop}
\tilde{W}_{\rm{b}}(i) =
\sum_{j=1}^{N_{\rm{b}}(i)}\exp\left(-\frac{z_{ij}^2}{\left( \alpha r_0 \right)^2}\right) ,
\end{equation}
with the parameter $\alpha$ chosen to be $\sigma/r_0$. This particular choice of the modulation function and of the associated parameter $\alpha$ allows us to identify both the dominant symmetry of the ordered structure, as well as more complicated, three-dimensional particle arrangements. The modified order parameters of the system with $N$ particles, $\Psi_m$, are then computed via

\begin{equation}
\label{eq:tot_boop}
\Psi_{m} = \frac{1}{N}\sum_{i=1}^{N} \left| \Psi_{m}(i)\right| .
\end{equation}

In our investigations, the neighbors of a tagged particles are identified with a three-dimensional Voronoi construction\cite{Voronoi:Book,Mickel:13,Steinhardt:83}, where the impenetrable walls are also considered as possible faces of the Voronoi polyhedra. The computation time required to perform three-dimensional Voronoi constructions can be prohibitively long as it scales with $N^3$. Even though the use of a neighbor list strongly enhances the efficiency of the computations, these improvements are still insufficient for systems with $N \gtrsim 100$. Therefore, the modified order parameters as defined in Equations (\ref{eq:local_boop}) - (\ref{eq:tot_boop}) have been computed only in our optimization calculations (see Subsection \ref{subsec:methods}) where the number of involved particles is smaller than 100 and only one single configuration has to be considered ({\it i.e.}, the one corresponding to the optimum particle arrangement).

\subsection{Methods}
\label{subsec:methods}

\subsubsection{Optimization strategies based on evolutionary algorithms}
\label{subsubsec:EA}

For a given slab width $H$, a particular value of the number density $\rho$, and at $T = 0$, the ordered equilibrium structure of a system is the configuration with the lowest energy ({\it i.e.}, lattice sum), corresponding to the global energy minimum of ``all possible ordered particle arrangements'' (to be specified below). A structure in a slab geometry that is periodic in the $x$- and $y$-directions and which contains $N$ particles per cell can be specified via a two-dimensional unit cell, spanned by two lattice vectors, $\bm{a}$ and $\bm{b}$, and the positions of the particles within this cell. In our formalism, we kept the base area of the unit cell $A$ constant, corresponding to a constant density $\rho$; thus the lattice vectors $\bm{a}$ and $\bm{b}$ can be parameterized as

\begin{equation}
\label{eq:lattice}
\displaystyle\bm{a}=\left(\begin{array}{c} \frac{A}{b_y} \\ 
0 \end{array}\right)\mbox{   };
\mbox{   }\displaystyle\bm{b}=\left(\begin{array}{c} b_x \\ 
b_y \end{array}\right) .
\end{equation}
Particle positions within the unit cell are specified by linear combinations of fractions of the lattice vectors and of the (effective) slab width $h$, respectively. Thus the parameter space that contains all possible particle arrangements is spanned by $(3N-2)$ parameters.

In order to identify the configuration with the lowest energy we employed an optimization tool that is based on ideas of EAs, using a phenotype implementation (for details {\it cf.} Refs. [\onlinecite{Gottwald:05}, \onlinecite{Deaven:95,Pauschenwein:08,Fornleitner:09}]). In addition to the standard operations used in EAs (such as of crossover- and mutation operations), candidate configurations are locally relaxed using the limited-memory Broyden-Fletcher-Goldfarb-Shanno algorithm \cite{Byrd:95} in its bounded variant (L-BFGS-B), which leads to a considerable speed-up in the convergence of the algorithm. To allow for a better parallelization of the code on multiple-CPU computers, we abandoned the traditional concept of generations and used a pool of ten individuals, instead \cite{Doppelbauer:12, Doppelbauer:12_a}.

For the creation of a new individual (which represents a particular ordered particle arrangement), two parents are chosen from this pool. The probability for an individual, ${\cal I}$, to be chosen as a parent is related to its fitness value, $f({\cal I})$; this value, in turn, depends on how well the selected individual is adapted to the problem. In our case, the fitness is related to the energy: individuals with lower energy are characterized by a higher fitness value. To be more specific, we have used in our work the following fitness function for individual ${\cal  I}$

\begin{equation}
\label{eq:fitness}
f({\cal I}) = 
\exp \left( -P_{\rm fit} \frac{E_{\cal I} - \rm{min}_{\{ \cal J \}}(E_{\cal J})} 
{\rm{max}_{\{ \cal J \}}(E_{\cal J}) - \rm{min}_{\{ \cal J \}}(E_{\cal J})} \right) .
\end{equation}
In this relation, $P_{\rm{fit}}$ is a numerical parameter that specifies to which degree individuals with higher fitness values are preferred in the selection process. ``min$_{\{ \cal J \}}(E_{\cal J})$'' and ``max$_{\{\cal J \}}(E_{\cal J})$'' are the minimum and maximum values of energy, respectively; they are taken over the set of all individuals, $\{ {\cal J} \}$. After a local relaxation, the fitness of the new individual is calculated; depending on this value, the individual is added to the pool or not, replacing in the former case a less suitable element of the pool.

For the repulsive IPL interactions considered in the present work, the energy landscape in search space does not have pronounced local minima; thus in our optimization scheme it was sufficient to create for each state point in total 500 individuals. In an effort to include even more complex structures that are characterized by a larger number of particles per unit cell, we have performed optimization runs with up to 28 particles per unit cell and have then compared the respective, optimized structures.

\subsubsection{Monte Carlo simulations}
\label{subsubsec:MC}

As will be discussed in detail in Section \ref{sec:results}, we have identified at {\it vanishing} temperature a cascade of complex intermediate structures that occur as the system transforms, with increasing slab width $H$, from a triangular to a square structure (or vice versa). In an effort to verify the mechanical stability of these phases at {\it finite} temperature, we have performed extensive MC simulations for a few select intermediate structures, {\it i.e.}, for given $H$-, $\rho$- and $T$-values. For these computations, we have introduced -- in addition to the system parameters already defined above --  the reduced temperature, $T^*$, via $T^*=k_{\rm{B}}T/4\epsilon$; for notational convenience, the asterisk will be dropped in the following. 

To be more specific, we have investigated the ground state configurations of four intermediate crystal phases (they will be denoted by I$_2$, I$_3$, I$_5$, and I$_6$) as they were identified within the optimization scheme based on EAs (see Fig.\ \ref{fig:i2} and Supplementary Material\cite{ref:supplementary}). The underlying lattices for these intermediate phases are rather complex and are often specified via non-orthorhombic cells; therefore we have used the particle configurations as predicted in the EA approach as initial configurations for the MC simulations, replicating the primitive cells in such a way that each initial configuration hosts at least 2000 particles in the simulation box. In addition, periodic boundary conditions and the minimum image convention are used. 

For all systems investigated via simulations, the initial configurations have been relaxed during $t_{\rm eq}$ MC cycles and averages have been accumulated over  $t_{\rm av}$ MC cycles, where $N$ trial moves are performed in one MC cycle. The amplitudes of the trial moves are chosen such that the acceptance ratio lies between 30\% and 60\%. In the Table we have summarized the most relevant input and output data of the MC runs for each of the four intermediate phases investigated.

To characterize the spatial order of the particles in the respective intermediate phases, we have computed the density profiles, $\rho(z)$, via

\begin{equation}
\label{eq:rhoz}
\rho(z) = \frac{1}{A}
\left \langle \sum_{i=1}^{N}\delta\left(z-z_i \right) \right \rangle
\end{equation}
with $\langle . \rangle$ being the ensemble average computed in the MC simulations and $A$ being the base area of the simulation cell. These density profiles are normalized via

\begin{equation}
\label{eq:Norm_rhoz}
\displaystyle \int_{-\frac{h}{2}}^{\frac{h}{2}} \rho(z)dz=\rho .
\end{equation}

In order to compute the in-plane, center-to-center correlation functions, $g_m(s)$, the slab is split into $M$ slices as follows (the index $m$ specifying the slice): we assume $\sigma/10$ for the width of the first ({\it i.e.}, bottom) and for the $M$-th ({\it i.e.}, top) slices, while we choose for the remaining $(M - 2)$ slices a width of $\Delta z= (h-\sigma/5)/(M-2)$. The vertical extent of slice $m$ is defined via the relation $z_m^{(-)}\leq z < z_m^{(+)}$, with 

\begin{equation}
\label{eq:Slices}
\left\{
\begin{tabular}{lllll}
$z_1^{(-)} = -\frac{h}{2}$ & ~~~;~~ & 
$z_1^{(+)} = -\frac{h}{2} + \frac{\sigma}{10}$ \\ 
$z_{M}^{(-)} = \frac{h}{2} - \frac{\sigma}{10}$ & ~~~;~~& 
$z_{M}^{(+)} = \frac{h}{2}$ \\ 
$z_{m}^{(-)} = z_1^{(+)} + (m-2) \Delta z$ & ~~~;~~ & 
$z_{m}^{(+)} = z_{m+1}^{(-)}$ & ~~~~~ & $1 < m < M$ .
\end{tabular}
\right.
\end{equation}

The number of particles, $N_{m}$, that are located in slice $m$ is then given by 
\begin{equation}
\label{eq:gm1}
N_{m} = \sum_{i=1}^{N} \int_{z_{m}^{(-)}}^{z_{m}^{(+)}}\delta(z-z_{i}) dz .
\end{equation}
To determine which slice a given particle belongs to, we define the distribution

\begin{equation}
\label{eq:gm2}
Y_{m}(z) = \Theta(z-z_{m}^{(-)})-\Theta(z-z_{m}^{(+)}) ,
\end{equation}
$\Theta(x)$ being the Heaviside function. With the above notations and definitions, the in-plane, center-to-center correlation function $g_{m}(s)$ for slice $m$ is defined as

\begin{equation}
\label{eq:gm3}
g_{m}(s) = A^2 \left \langle \frac{1}{N_m^2}
\sum_{i=1}^{N}\sum_{i\neq j}\delta(\bm{s}-\bm{s}_{ij})
Y_m(z_i) Y_m(z_j) \right \rangle ,
\end{equation}
with ${\bf s}_{ij} = {\bf s}_i - {\bf s}_j$; here ${\bf s}_i$ and ${\bf s}_j$ are two-dimensional vectors, obtained via projections of the corresponding three-dimensional position vectors, ${\bf r}_i$ and ${\bf r}_j$, of particles $i$ and $j$ onto the $(x, y)$-plane. In this expression, the distribution $Y_m(z)$ selects particles $i$ and $j$ in the slice $m$, while the factor $\delta({\bf s} - {\bf s}_{ij})$ projects the positions of the particles onto a plane parallel to the $(x, y)$-plane. With these definitions at hand, the correlation functions satisfy $g_{m}(s \rightarrow \infty) \simeq 1$.

For each of the intermediate phases, the functions $\rho(z)$ and $g_{m}(s)$ show particular features that permit to verify qualitatively the mechanical stability of the phases.

The consistency between the results obtained for the energy via the EA approach and the MC simulations has been investigated by using the multiple histogram reweighting method \cite{Ferrenberg:88,Ferrenberg:89,Newman:book:99} for the intermediate phases I$_2$ and I$_3$. Extrapolating the average energy per particle obtained at {\it finite} temperatures towards {\it zero} temperature (corresponding to the ground state energy) allows us to verify the consistency between the two different approaches.

\section{Results}
\label{sec:results}

\subsection{Ground state energies and structures}
\label{subsec:ground_state}

\begin{figure*}
\begin{center}
\includegraphics[width=14cm]{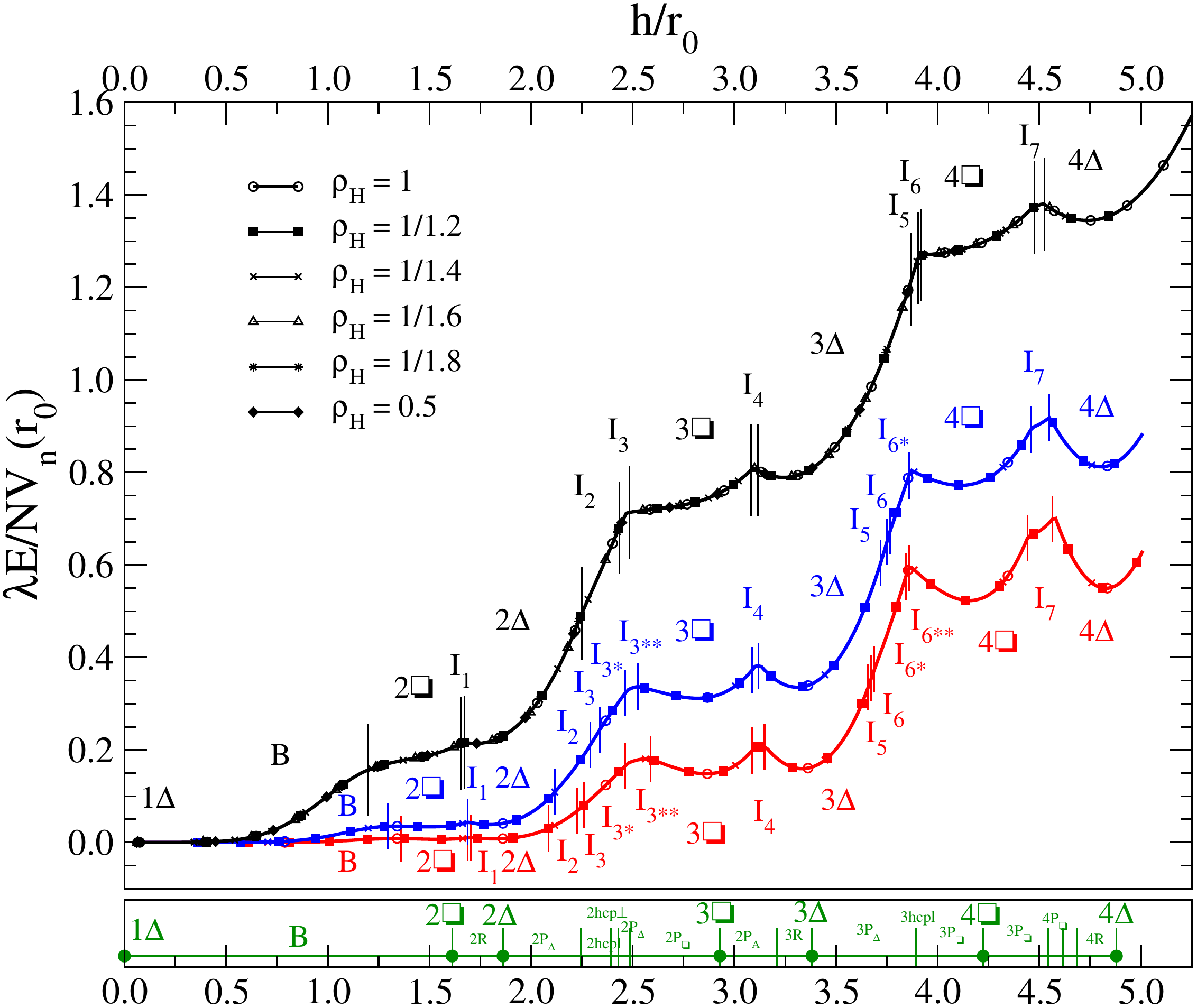}
\end{center}
\caption{\label{fig:density} Scaled global energy minima, $\lambda E/N V_n(r_0)$, computed with the EA-based optimization tool, as functions of the ratio $h/r_0$ for three select values of softness, $s=1/n$, and for select values of the bulk densities $\rho_H$ (as labeled). The number density is given by $\rho=(1+\sigma/h)\rho_H$; results obtained for different $\rho$-values are specified by different symbols (as labeled). Further, $r_0=(3/4\pi\rho)^{1/3}$ and $V_n(r_0)=4\epsilon(\sigma/r_0)^n$. The curves display data for $n=12$ (black line), $n=21$ (blue line) and $n=30$ (red line). Labels along the curves specify ordered structures (see text); their respective ranges of stability are delimited by thin, vertical lines. Note that -- in an effort to present all the data on a single vertical scale -- the values of $E/NV_n$ have been re-scaled by suitably chosen factors $\lambda$ ($\lambda=10^{3}$ for $n=12$, $\lambda=4\times 10^{5}$ for $n = 21$, and $\lambda=10^{8}$ for $n= 30$). The different structures are specified by labels that are introduced in the text; `B' stands for the buckled phase\cite{Oguz:12} (not discussed in the text). The bottom bar (green) shows the corresponding phase diagram of hard spheres (equivalent to $n\to \infty$; {\it cf.} Refs.\ [\onlinecite{Oguz:09},\onlinecite{Oguz:12}]), rescaled to the same units. The green dots indicate the values of $h$ where square or triangular multi-layer systems are stable.}
\end{figure*}

We have started our investigations by identifying the ordered ground state structures ({\it i.e.}, at $T=0$) for our system; this was achieved by applying the optimization scheme outlined in subsection \ref{subsec:methods}. We have chosen three values for the exponent $n$, namely $n = 12$, 21, and 30. Investigations were carried out for six different values of $\rho_H$ (specified in the inset of Figure \ref{fig:density}), varying the (effective) slab width $h$ over a representative range ($0<h \leq 5 r_0$).  For $h = 0$, the slab is a monolayer for which the ground state is a triangular lattice, while for $h \to \infty$ we recover the three dimensional bulk system for which, depending on the softness parameter\cite{Agrawal:95} $1/n$, the ground state is either bcc (for $1/n>1/6.25$) or fcc (for $1/n<1/6.25$). For each combination of these three system parameters, we have considered up to 28 particles per unit cell; the respective ground state structures were recorded. From those particle arrangements the one with the lowest energy, computed as a lattice sum, was retained as the respective ground state for the system. We have varied $h$ in the range $[0, 5 r_0]$, assuming an increment of $\Delta h = 0.01 \sigma$.

At finite temperature and for a three-dimensional bulk geometry (recovered in the limit $h \to \infty$), the properties of our system depend on a single parameter, namely the coupling parameter $\Gamma$, defined via 

$$
\Gamma = \frac{4 \epsilon}{k_{\rm B} T} \left( \frac{4 \pi \rho \sigma^3}{3} 
\right)^{n/3} ;
$$
for a fixed value of $n$, a single isotherm or isochore is therefore sufficient to describe the entire phase diagram of the system. As the system is confined in a slab geometry ({\it i.e.}, for finite $h$), the thermodynamic properties of the system depend now on two parameters, namely $\Gamma$ and $h/r_0$. The scaling of the energy shown in Figure \ref{fig:density}  is therefore a direct consequence of the scaling properties of the IPL interaction. Thus, by applying a suitable scaling prescription for both the energy, $E$, as well as for the slab width, $h$, the energy curves obtained for one particular value of $n$ can be made to collapse onto one single master curve, irrespective of the value of the density $\rho$. To be more specific, $E$ has to be divided by $V_n(r_0)$ [{\it cf.} Equation (\ref{eq:IPL})] and the slab width $h$ has to be divided by $r_0$, thereby scaling out the density $\rho$. Because of the scaling properties of IPL interactions at a given slab width $h$, the dependence of the global energy minimum on the density $\rho$ is completely included in $V_n(r_0)$ and the energy curves obtained for different values of $\rho$ collapse, for a given $n$-value, onto a single master curve. The resulting curves, {\it i.e.}, $E / V_n(r_0)$ {\it vs.} $h/r_0$ are displayed for the three $n$-values considered in this study in Figure \ref{fig:density}. In an effort to improve the visibility of the data, each curve has been multiplied by a suitably chosen pre-factor $\lambda$. The different symbols, as specified in the labels of Figure \ref{fig:density}, represent different values of density.

\begin{figure}
\begin{center}
\includegraphics[width=7cm]{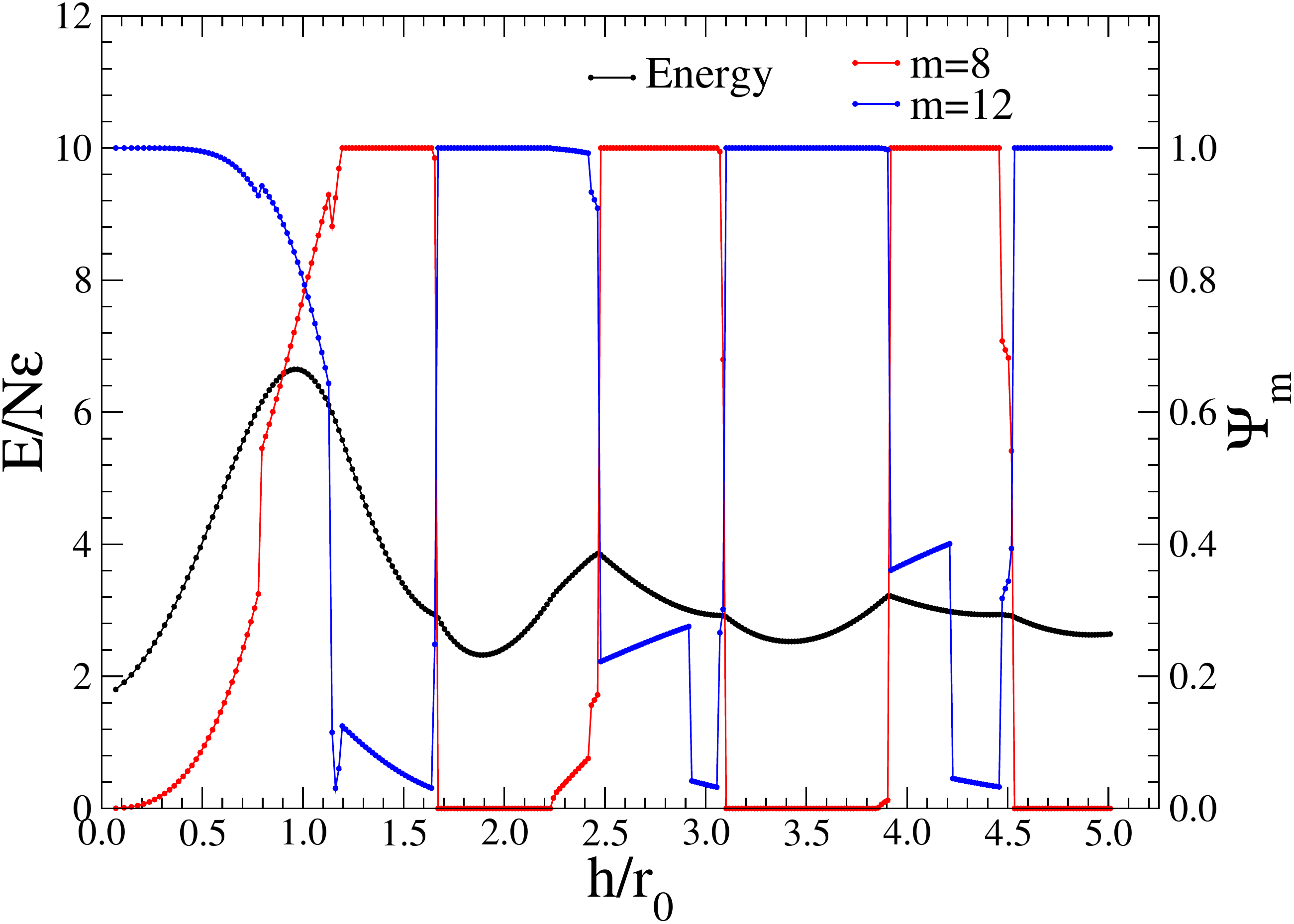}
\end{center}
\caption{\label{fig:order} Energy per particle, $E/N\epsilon$, of the ordered ground states for the system investigated in this contribution as a function of $h/r_0$, assuming $n=12$ and $\rho_H=1/1.2$ (black line). The order parameters $\Psi_8$ (red line) and $\Psi_{12}$ (blue line) indicate both the occurrence of alternating square and triangular phases as well as the identified intermediate phases.}
\end{figure}

To characterize these structures, the order parameters $\Psi_8$ and $\Psi_{12}$, as defined in Equations (\ref{eq:local_boop}) - (\ref{eq:tot_boop}) are used. For $n = 12$ and for $\rho_H = 1/1.2$, the corresponding values for $\Psi_8$ and $\Psi_{12}$ are reported in Figure \ref{fig:order}, along with the ground state energy $E$ as functions of $h/r_0$. When $\Psi_8 = 1$ or $\Psi_{12} = 1$, the ground state has a square or triangular symmetry, respectively; as mentioned above, we denote these lattices by symbols $\Box$ and $\triangle$ as defined for colloidal systems\cite{Pieranski:80,Pieranski:83}. 

A closer, quantitative analysis of our structures reveals that over large portions of the considered $h$-range, the systems form simple multi-layer structures with either square or triangular symmetry, denoted by $n_l \Box$ or $n_l \triangle$, respectively; $n_l$ being the number of layers\cite{Pieranski:80,Pieranski:83}. This sequence of structures, which is common to all three $n$-values investigated, can thus be formally written as $1 \triangle \to 2 \Box \to 2 \triangle \dots$~. In addition to these rather simple particle arrangements, we have identified structures which occur as the system either transforms with increasing $h$ from a square to a hexagonal structure ({\it i.e.}, $n_l \Box \to n_l \triangle$) or as it creates a new layer [{\it i.e.}, $n_l \triangle \to (n_l + 1) \Box$]. These intermediate structures are denoted in a rather arbitrary way by I$_k$ with $k$ being an integer. Additional structures which were observed for an exponent $n$ other than 12 are indicated by asterisks. A first look at Figure \ref{fig:density} reveals that in general these intermediate structures represent ground state energies in rather small intervals in the variable $h$; often they are only related to a single point in our $h$-grid (see specification of $\Delta h$ above). Therefore we cannot exclude the possibility that additional intermediate structures could emerge as ground state lattices in even smaller intervals than the ones we could grasp with our $h$-grid.

At the bottom of Figure \ref{fig:density} we have added for reference the sequence of ordered structures that were obtained in essentially the same setup for a system of {\it hard spheres}\cite{Oguz:12}, {\it i.e.}, a system which is formally obtained by making the exponent in our pair potential [{\it cf.} Equation \ref{eq:IPL}] tend towards infinity. Thus, the sequence of results obtained for $n = 12$, $n = 21$ and $n = 30$ should 'tend' towards the results summarized in the horizontal diagram at the bottom of Figure \ref{fig:density}. Note that for a hard sphere system the situation is -- as compared to soft spheres -- different in the sense that the stability regions of multilayer structures (such as $n_l \triangle$ or $n_l \Box$) reduce to isolated points on the $h$-axis: as already addressed in the Introduction, these $h$- (or $H$-values) correspond to the thickness of $n_l$ slides of the fcc bulk crystal, {\it i.e.}, the structure is the result of $n_l$ triangular or square stacked layers.

The $h$-ranges where the layered structures with square or triangular symmetry ({\it i.e.}, $n_l \Box$ or $n_l \triangle$) are stable can easily be identified in Figure \ref{fig:order} as those segments where the order parameters $\Psi_8$ and $\Psi_{12}$ are either 0 or 1. Abrupt changes in these order parameters at the respective edges of these segments and intermediate values of these parameters indicate the occurrence of an intermediate phase, I$_k$. Two of these ground state particle arrangements are discussed in the following.

\begin{figure*}
\includegraphics[width=5cm]{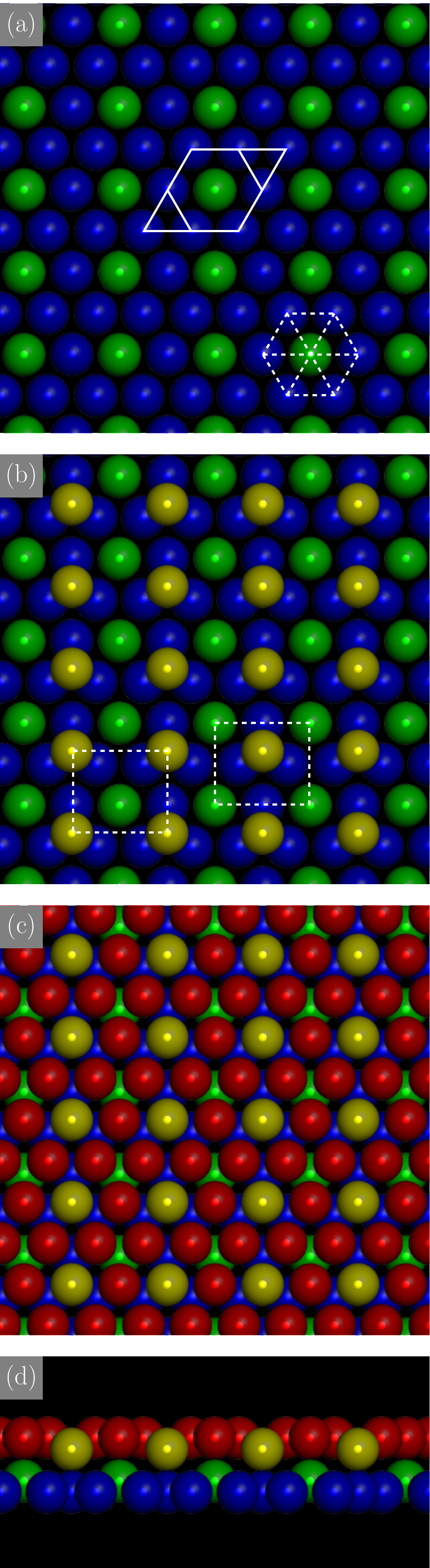}
\hspace{3.5cm}
\includegraphics[width=5cm]{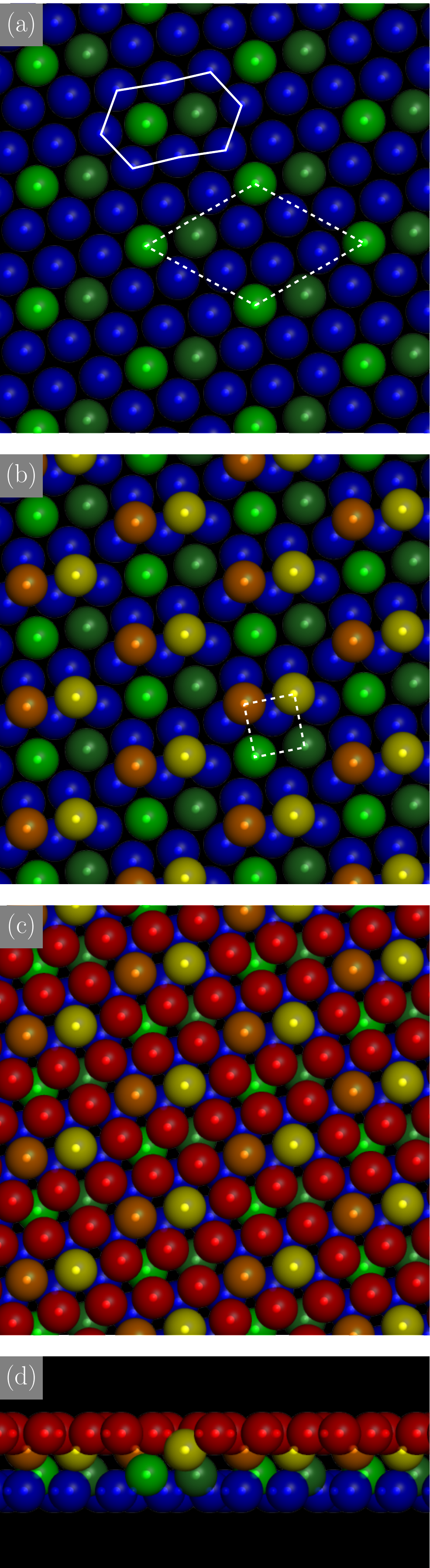}
\caption{\label{fig:i2} Intermediate structures identified for $n=12$ and $\rho_H=1/1.2$. Left panels show structure I$_2$ at $h=1.26\sigma$, right panels show structure I$_3$ at $h=1.34\sigma$. Panels (a) and (b): top views showing the layer-by-layer growth of the respective structure; thus, panels (c) show the the top views of the respective structure. Panel (d): side views of the respective intermediate structure. Particles are colored according to the layer they belong to: blue -- bottom layer, dark green/green -- second layer, yellow/orange -- third layer, and red -- top layer. White lines indicate structures discussed in the text.}
\end{figure*}

Details of the intermediate structure I$_2$ are depicted in the left half of Figure \ref{fig:i2}: two of the panels [labeled (a) and (b)] visualize the growth of the structure layer-by-layer, providing thereby a better understanding of the complexity of the internal architecture of this four-layer structure; further, panels (c) and (d) show the {\it full} structure I$_2$ in two orthogonal views. For convenience, particles belonging to the respective layers are shown in different colors. The building blocks of the bottom layer of this structure is a regular, hexagonal ring of particles  (colored blue) with two adjacent, oppositely oriented equilateral triangles; these units are staggered in reverse order. Particles of the subsequent layer (colored green) are elevated by a moderate vertical displacement (that can be estimated from the side view shown in panel (d) of Figure \ref{fig:i2}) above the centers of the hexagonal rings of the bottom layer, forming thereby in a second layer a rectangular lattice. Thus these particles can be viewed as the tips of flat, six-sided pyramids whose bases are formed by the hexagons of the bottom layer. Particles belonging to the third layer (colored yellow) form again a rectangular lattice, which is identical to the one formed by the particles of the second layer; with respect to the latter, it is shifted both vertically and horizontally. Again, these particles can be viewed as the tips of flat, six-sided pyramids, which are now oriented upside down; the basis particles (colored red) of these pyramids finally form the fourth and thus the top layer of the intermediate structure I$_2$. A structure qualitatively similar to I$_2$ has also been observed in confined Yukawa systems at appropriate values for the screening length\cite{Oguz:09}.

The intermediate structure I$_3$ is an even more complex four-layer structure; it is depicted in the right half of Figure \ref{fig:i2}: two of the panels [labeled (a) and (b)] visualize the growth of the structure layer-by-layer, providing thereby a better understanding of the complexity of the internal architecture of this four-layer structure; further, panels (c) and (d) show the {\it full} structure I$_3$ in two orthogonal views. For convenience, particles belonging to the respective layers are shown in different colors. The basic units of the bottom and of the top layers (particles colored blue and red, respectively) are now elongated rings formed by eight particles. The central areas of the bottom rings are thus able to accommodate two particles of the subsequent layer (colored dark green/green), which is again moderately elevated in the vertical direction with respect to the bottom layer. The third and the fourth layer (particles colored yellow/orange and red, respectively) are -- similar as in the intermediate structure I$_2$ -- mirror images of the two bottom layers, suitably shifted in the horizontal direction; we note that these layers are not perfect in the sense that particles belonging to this layer differ within $1.2\%$ in their vertical positions: in a projection orthogonal to the layers, the particles of the second and of the third layers form together a square lattice; with their positions, the particles of the third layer fix the positions of the particles of the top layer which again form the aforementioned complex arrangements of elongated eight particle rings.

As mentioned before, all ground state configurations that we have identified in our investigations are summarized in the Supplementary Material\cite{ref:supplementary}. At this point it is appropriate to briefly establish a connection between these particle arrangements and other ordered configurations between two and three dimensions, identified in previous scientific contributions for particles with similar interactions\cite{Oguz:09,Oguz:12,Schmidt:97,Zangi:00a,Zangi:00b}; in this context we also refer the interested reader to the review article Ref. \onlinecite{Loewen:12}.
The particle arrangement that we have denoted I$_1$ corresponds to structure 2R, identified in Ref. \onlinecite{Oguz:12,Schmidt:97} for hard spheres. Our particle arrangement I$_2$ corresponds to the ''Belgian waffle iron'' lattice observed in Ref. \onlinecite{Oguz:09} for a confined system of Yukawa particles; in contrast, for our structure I$_3$ we could not establish a link to previously identified lattices, so far. Structures I$^*_3$ and I$^{**}_3$, identified in the present contribution correspond to lattices formed by hard sphere particles that were denoted in Ref. \onlinecite{Oguz:09,Oguz:12} as 2P$_\triangle$ and 2P$_\Box$ structures, respectively. Structure I$_4$ was previously identified for hard spheres in Ref. \onlinecite{Oguz:12} as lattice 3R. Structures I$_5$ and I$_6$ have -- to the best of our knowledge -- not been identified, so far. Particle arrangements I$^*_6$ and I$^{**}_6$ corresponds most likely to structures 3P$_\Box$ and 3P$_\triangle$, respectively, identified for hard spheres in Ref. \onlinecite{Oguz:12}. Finally, the particle arrangement I$_7$ carries traces of the structures 4R, 4P$_\Box$ and 3P$_\Box^l$, reported for hard spheres in \onlinecite{Oguz:12}.

\subsection{Monte Carlo simulations}
\label{subsec:MC}

Using extensive MC simulations we have investigated the energies and the structural properties of select intermediate particle arrangements at {\it finite} temperatures. In order to be able to calculate these properties with an acceptable computational effort over a representative range of temperatures, we have used the multiple histogram reweighting method, proposed by Ferrenberg and Swendsen\cite{Ferrenberg:88,Ferrenberg:89}. In this approach, MC simulations are carried out for a few, select temperatures, evaluating thereby the energy histograms. Based on these results, the histograms for neighboring temperatures can be obtained via inter- and extrapolation. 
 
In this Subsection we report on our investigations of the properties of the intermediate structures I$_2$, I$_3$, I$_5$, and I$_6$. The Table contains the most relevant input and output data of the respective simulations.

\begin{figure}
\begin{center}
\includegraphics[width=7cm]{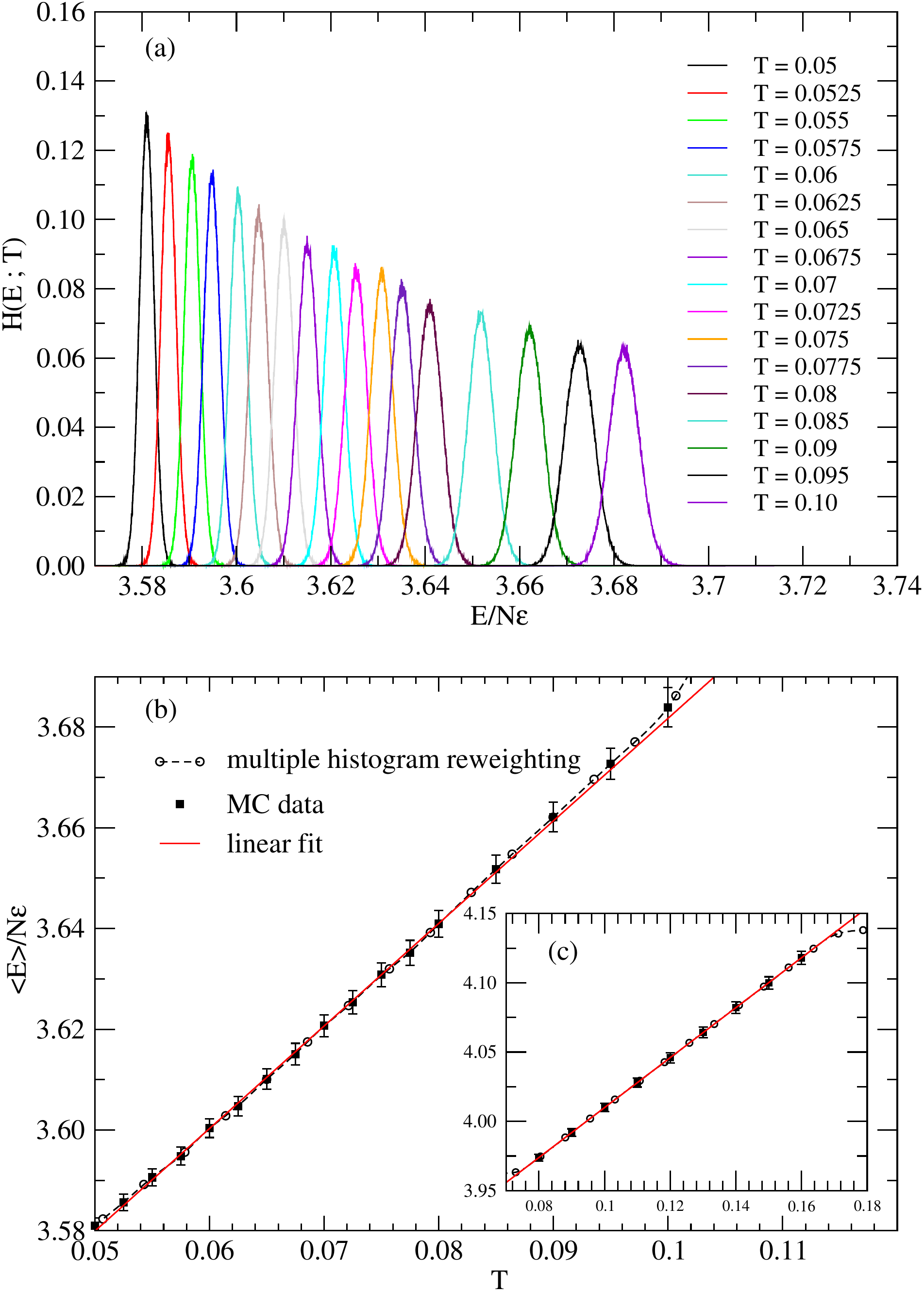}
\end{center}
\caption{\label{fig:histograms} Panel (a): Energy histograms, $H(E, T)$, evaluated for the intermediate structure I$_2$ for temperatures as labeled for $n=12$ and $\rho_H=1/1.2$; histograms have been calculated via the multiple histogram reweighting approach \cite{Ferrenberg:88,Ferrenberg:89}. Panel (b): Average energy per particle, as evaluated in MC simulations via the multiple histogram reweighting method for the intermediate phase I$_2$ (symbols as labeled). Panel (c): same as panel (b), but for the intermediate phase I$_3$.}
\end{figure}

In panel (a) of Figure \ref{fig:histograms} we display these energy histograms, $H(E; T)$, evaluated over a representative temperature range ({\it i.e.}, $0.05 \le T \le 0.1$) for the intermediate structure I$_2$. The average values of energy per particle, $\langle E \rangle /N\epsilon$, are plotted as functions of the temperature in panels (b) and (c) of this Figure for the intermediate structures I$_2$ and I$_3$, respectively: in both cases, these data show over a remarkably broad temperature range a linear dependence on temperature; from a least-square fit of these data we obtain  

\begin{eqnarray} \nonumber
\left\{
\begin{tabular}{l}
$\langle E \rangle/N\epsilon  =  3.48 + 2.04 ~ T ~~~~ {\rm for~structure~I}_2$ \\ \nonumber
$\langle E \rangle/N\epsilon  =  3.83 + 1.80 ~ T ~~~~ {\rm for~structure~I}_3$ .
\end{tabular}
\right.
\end{eqnarray}

Via these relations we can easily extrapolate the corresponding energy values at $T = 0$, which thus correspond to the ground state energies; the data show a remarkable agreement with the energy values predicted by the optimization scheme: $E_0/N\epsilon = 3.4835298$ (for structure I$_2$) and $E_0/N\epsilon =3.8271842$ (for structure I$_3$). 

\begin{figure*}
\begin{tabular}{cc}
\includegraphics[width=7cm]{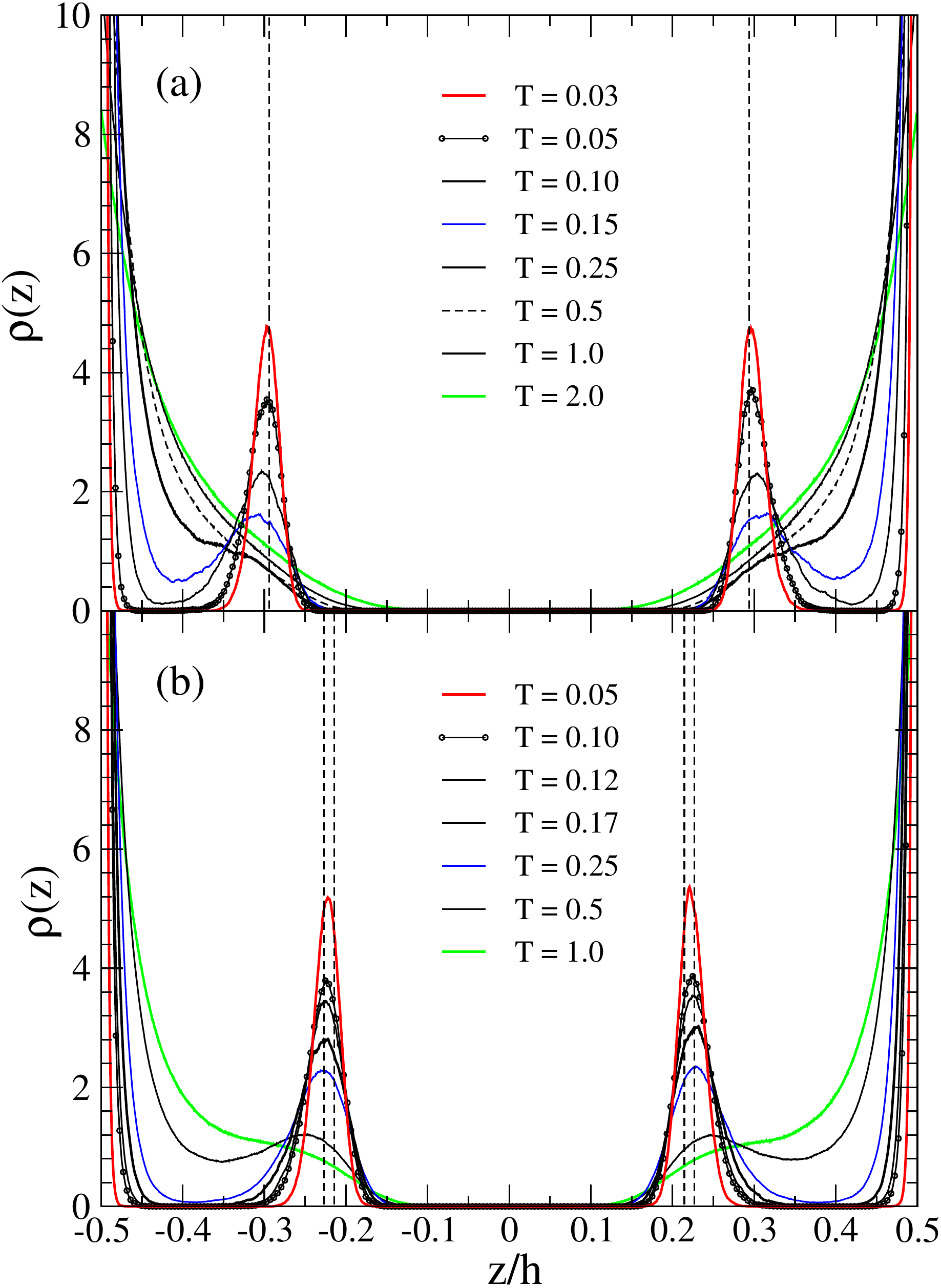} & 
\includegraphics[width=7cm]{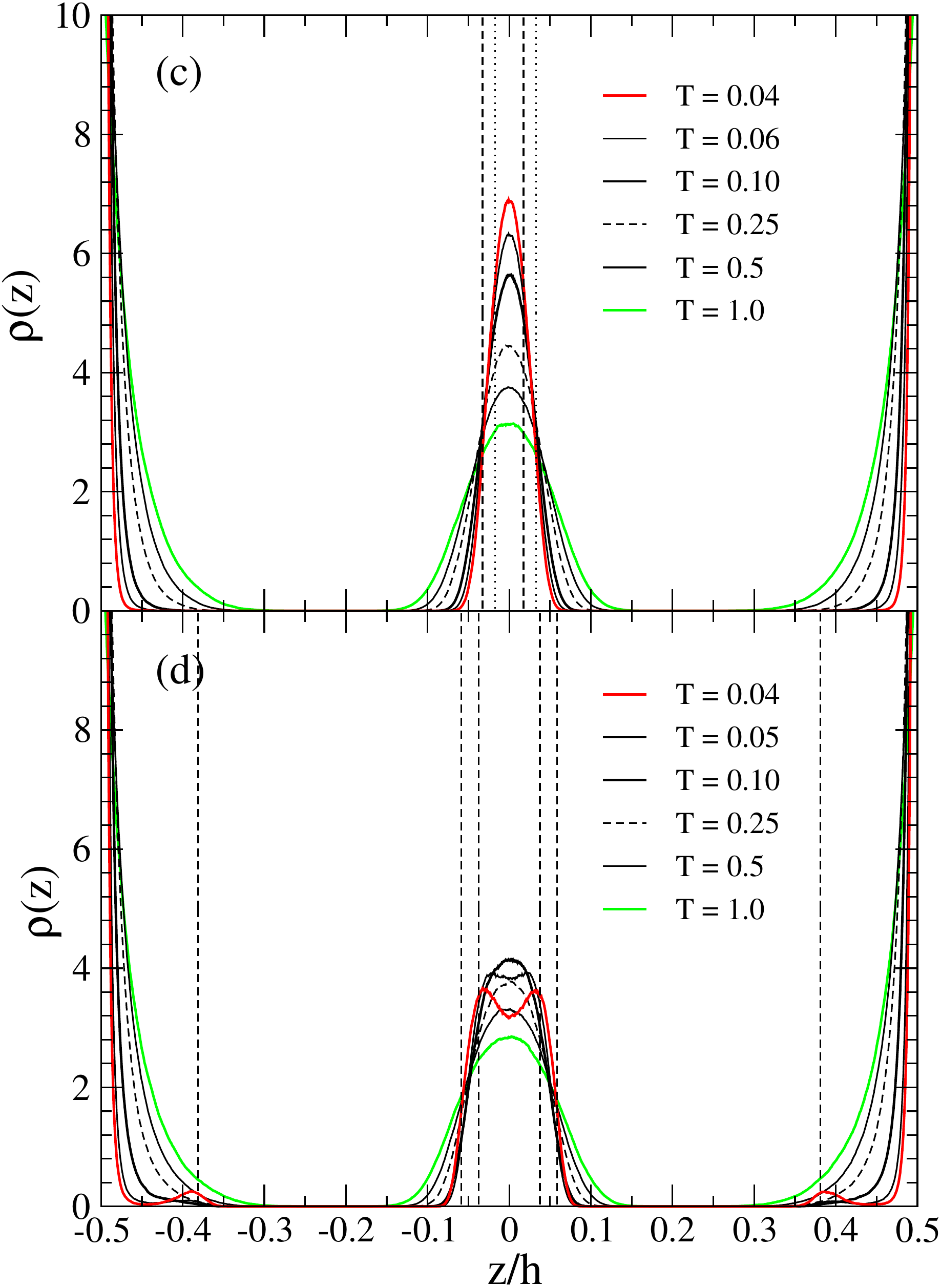}
\end{tabular}
\caption{\label{fig:density_profiles} Density profiles computed in MC simulations for some intermediate phases at different temperatures (as indicated) for $n=12$ and $\rho_H=1/1.2$. (a) $h=1.26 \sigma$ (I$_2$); (b) $h=1.34 \sigma$ (I$_3$); (c) $h=2.25 \sigma$ (I$_5$ or $3\triangle$, see text) and (d) $h=2.28 \sigma$ (I$_6$). Dashed and dotted vertical lines indicate the positions of the inner layers as predicted by the optimization scheme.}
\end{figure*}

In Figure \ref{fig:density_profiles} we show the vertical density profiles, $\rho(z)$, as defined in Equation (\ref{eq:rhoz}) for the intermediate particle configurations I$_2$, I$_3$, I$_5$, and I$_6$. The data were obtained for select values of the (effective) slab width, $h$, and for several temperatures (as specified in the caption and in the panels of Figure \ref{fig:density_profiles}). By tracing the peak heights of these density profiles, we can identify the respective temperatures, $T_{\rm s}$, above which the intermediate structures become mechanically unstable.

\begin{figure*}
\begin{center}
\includegraphics[width=14cm]{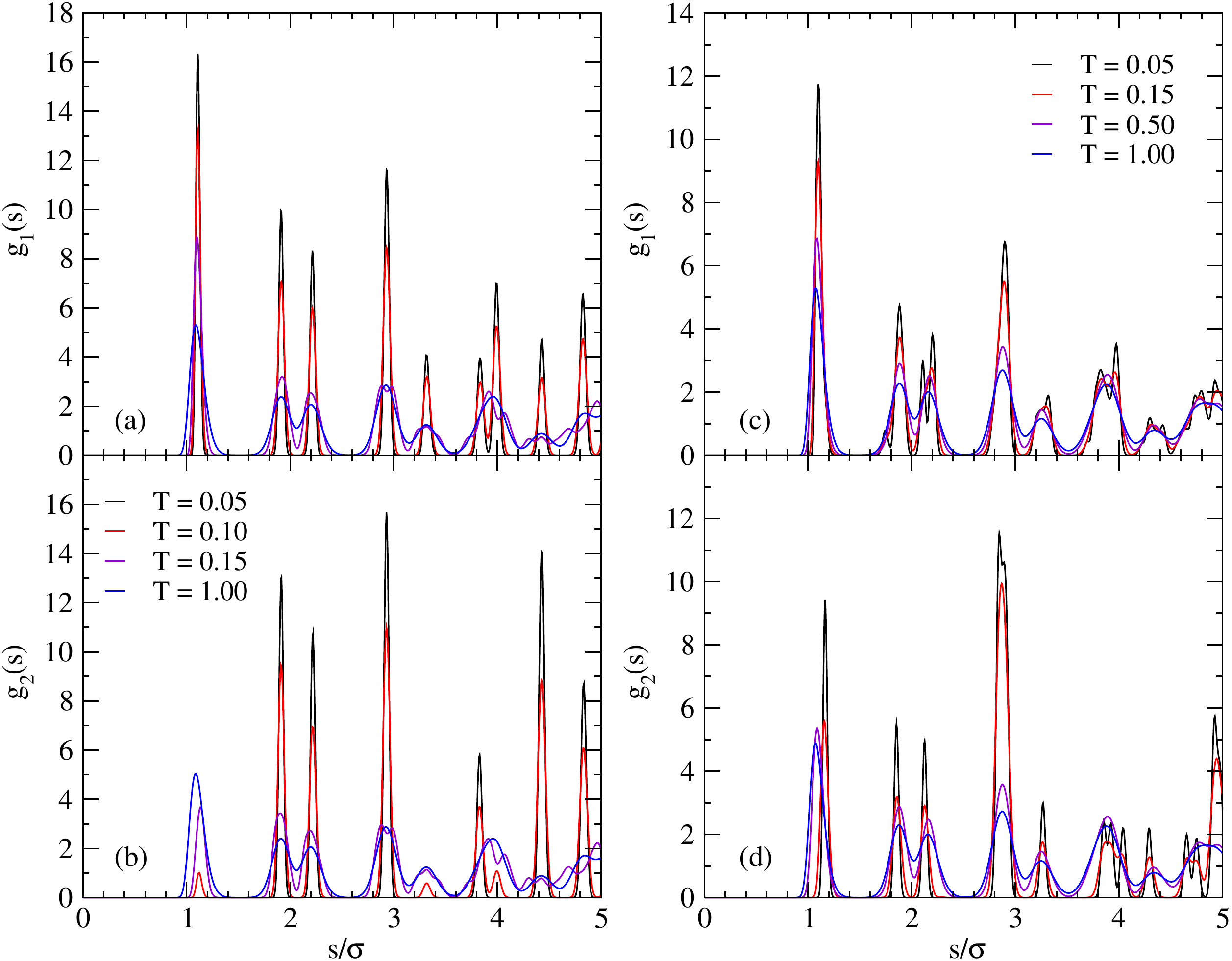}
\end{center}
\caption{\label{fig:correlation_functions} In-plane correlation functions $g_1(s)$ and $g_2(s)$ [as defined in Equations (\ref{eq:gm1}) - (\ref{eq:gm3})] for the intermediate phases I$_2$ [panels (a) and (b)] and I$_3$ [panels (c) and (d)] computed in MC simulations for $n=12$, $\rho_H=1/1.2$, and temperatures as labeled. The following values for the (effective) slab width, $h$, have been assumed: $h = 1.26 \sigma$ (I$_2$) and $h = 1.34 \sigma$ (I$_3$).}
\end{figure*}

The density profile of the intermediate structure I$_2$, evaluated for $h = 1.26\sigma$, is shown in panel (a) of Figure \ref{fig:density_profiles}; as discussed in Subsection \ref{subsec:ground_state}, I$_2$ is a four-layer structure, thus $M = 4$; {\it cf.} Equation (\ref{eq:gm3}). The positions of the two pronounced inner peaks (which survive up to $T_{\rm s} \simeq 0.15$) agree with high accuracy with the corresponding results obtained via the optimization scheme. For $T \gtrsim T_{\rm s}$, the intermediate structure becomes mechanically unstable and the two inner layers vanish [reflected by the fading of the two respective peaks in $\rho(z)$] -- {\it cf.} Figure \ref{fig:density_profiles}(a). At this temperature the intermediate structure I$_2$ transforms into a 2$\triangle$ structure where the particles fluctuate to a considerable amount in the vertical direction due to the elevated temperature. The sizable spatial extent of the two layers is nicely reflected (i) by the fact that the density profile $\rho(z)$ extends for $T\gtrsim 0.25$ for each layer over at least one third of the slab width (see panel (a) of Figure \ref{fig:density_profiles}) and (ii) by the observation that with increasing temperature the first peak in $g_1(s)$ decreases while the first peak in $g_2(s)$ increases, reflecting the tendency towards an increased vertical mobility of the particles (see panels (a) and (b) in Figure \ref{fig:correlation_functions}). The transformation from the structure I$_2$ to the 2$\triangle$ phase can also nicely be quantified via the in-plane correlation functions. While particles in the first slice arrange for all temperatures in a triangular lattice [reflected in the characteristic peak positions in $g_1(s)$], we observe a shift of the first peak in $g_2(s)$ with increasing temperature such that eventually, as the transition temperature $T_{\rm s}$ is reached, the positions of the peaks in $g_2(s)$ provide evidence of a triangular particle arrangement (see also a more quantitative discussion below).

\begin{figure}
\begin{center}
\includegraphics[width=7cm]{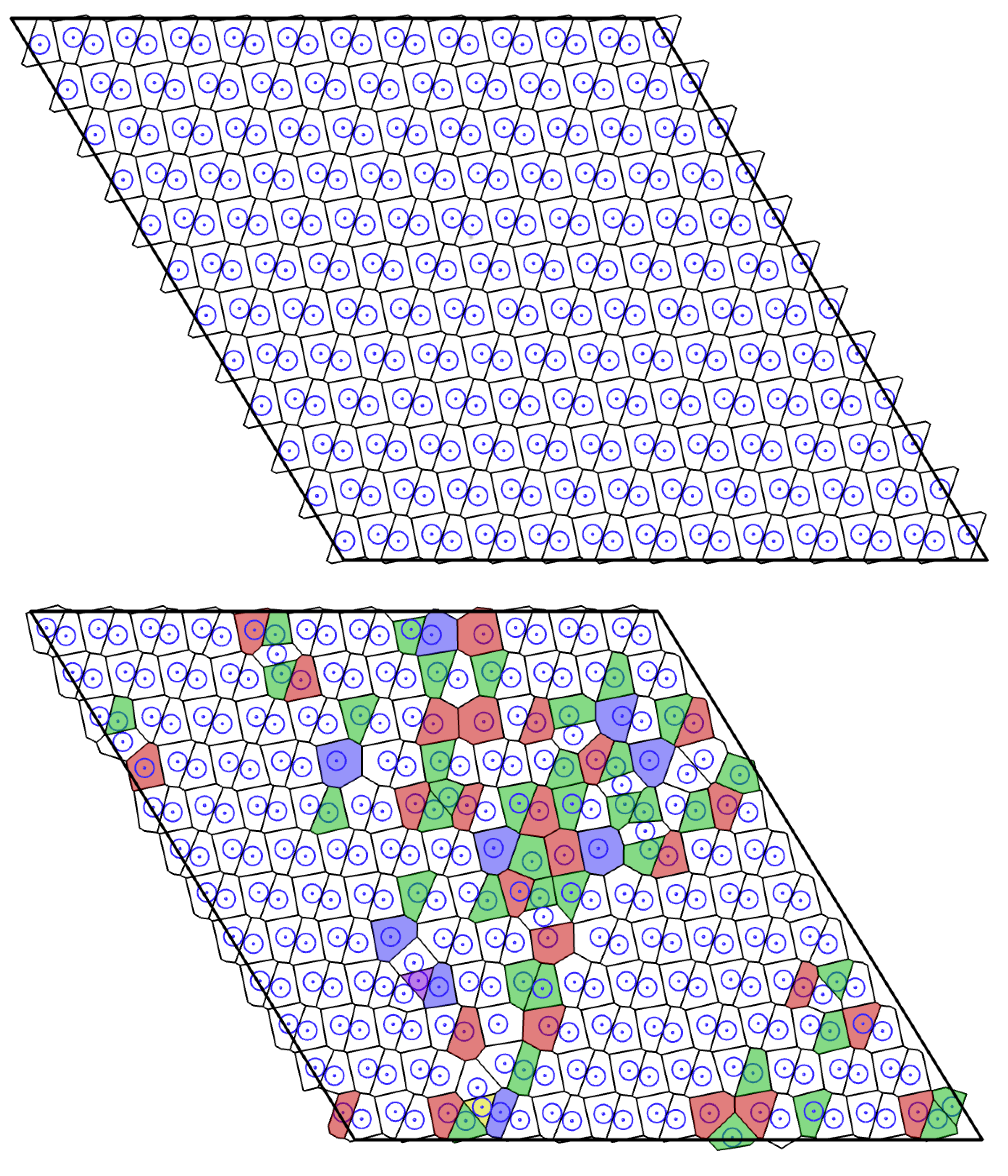}
\end{center}
\caption{\label{fig:snapshots} Snapshots of the MC simulation for $n=12$, $\rho_H=1/1.2$, $h=1.34 \sigma$, and $T=0.25$. Top panel: simulation cell for the second layer of the intermediate phase I$_3$ as used as an initial configuration for the MC simulations; the structure is composed of 144 primitive cells (with two particles per cell) as they were predicted by the EA-based optimization scheme. Lines delimit two dimensional Voronoi cells for this layer, the diameter of the circles is $\sigma$. Bottom panel: snapshot of the second layer of the intermediate phase I$_3$ obtained in MC simulations at {\it finite} temperature after $2.4 \times 10^6$ MC cycles; the chosen temperature corresponds to the limit of mechanical stability of phase I$_3$ ({\it i.e.}, $T = 0.25$). Again, lines delimit two dimensional Voronoi cells; the following color coding is used to provide information on the number of edges of these cells: three -- purple, four -- yellow, five -- green, six -- white, seven -- red, and eight -- blue.}
\end{figure}

In a similar analysis carried out for the intermediate structure I$_3$ (which is again a four-layer ground state configuration; thus $M = 4$), we conclude from the temperature dependence of the vertical density profile (displayed in panel (b) of Figure \ref{fig:density_profiles}) that this particle arrangement is stable for temperatures up to $T_{\rm s} \simeq 0.25$. This melting process can be visualized very nicely via an in-plane Voronoi construction carried out for $h = 1.34\sigma$ for the second slice of the intermediate phase I$_3$, shown in Figure \ref{fig:snapshots}; in accordance with the definitions of the slice boundaries specified in Equation (\ref{eq:Slices}), the vertical range of this slice is given by $-0.57\sigma\leq z \leq 0$. Starting from an initial configuration (shown in the top panel of Figure \ref{fig:snapshots}) that is imposed by the EA-predicted, ordered structure, many defects are generated at {\it finite} temperature ({\it i.e.}, $T = 0.25$) in the MC simulations; a typical particle arrangement for this temperature is shown in the bottom panel of Figure \ref{fig:snapshots}. At even higher temperatures, {\it i.e.}, for $T > 0.25$, the typical pair-arrangement of the particles does not persist; instead larger, linear clusters with ramifications are formed: a rough idea of these arrangements can be grasped from the bottom panel of Figure \ref{fig:snapshots}, in particular from those particles with three or four neighbors. As can also be seen in the density profiles of phase I$_3$ [{\it cf.} Figure \ref{fig:density_profiles}(b)], particles in the second layer of I$_3$ are not confined to a plane, which is similar to what happens at finite temperatures.

We now proceed to the intermediate phase I$_5$. Here, the identification of the limits of stability via MC simulations is considerably more difficult: as we can see in Figure \ref{fig:density}, this intermediate phase is a direct neighbor of structure 3$\triangle$ as we increase $h$. The characteristic difference between the structures I$_5$ and 3$\triangle$ lies in the vertical position of the particles close to $z=0$ (see Figure S7 in the Supplementary Material\cite{ref:supplementary}). At vanishing temperature the two inner layers of the intermediate structure I$_5$ are separated by a minute vertical distance (to be quantitative: they are located at $z_{\rm{I}_5}=\pm 0.112 \sigma$). Projecting the horizontal positions of the particles of both layers onto the plane $z=0$, they form a triangular lattice. At finite temperatures, particles fluctuate in their positions due to thermal agitations. Their vertical displacements can now easily extend over distances larger than $z_{\rm{I}_5}$, making it thus essentially impossible to distinguish at finite temperature in our MC data for the density profile $\rho(z)$ between a single layer (indicating the observation of a 3$\triangle$ structure) or two layers in close proximity (characterizing the intermediate I$_5$ structure). Proceeding to lower temperatures, where the resolution of a double peak might be possible, does not represent a successful alternative, since the acceptance ratio for particle moves in the MC simulation drops dramatically. Decreasing under these conditions the step size of the trial move would allow us to restore this ratio to convenient values, but then a proper sampling of the phase space becomes very inefficient and the correct exploration of phase space would require an unreasonable amount of CPU time.

In contrast, the particular features of phase I$_6$ are easier to identify via MC simulations. The central maximum in the density profile (shown in panel (d) of Figure \ref{fig:density_profiles}) is double peaked at low temperatures; in addition, two additional peaks are visible for $z\simeq \pm0.38\sigma$, indicating that a small number of particles escape from the top and the bottom layers. From our MC results we estimate that the limit of mechanical stability of the intermediate phase I$_6$ occurs at $T_{\rm s} \simeq 0.04-0.05$.

Finally, we discuss our results obtained for the in-plane correlation functions, $g_m(s)$, as defined in Equations (\ref{eq:gm1}) - (\ref{eq:gm3}); these functions also provide information about the stability of the intermediate phases. We report that in all our MC simulations the correlation functions are -- within numerical accuracy -- symmetric with respect to the $(z = 0)$-plane, {\it i.e.}, $g_1(s) = g_M(s)$ and $g_m(s) = g_{M+1-m}(s)$ for $1 < m \leq M/2$. Representative results for the intermediate phases I$_2$ and I$_3$ are shown in Figure \ref{fig:correlation_functions}: we display the in-plane correlation functions $g_1(s)$ and $g_2(s)$ for the intermediate structures I$_2$ ($h = 1.26\sigma$) and I$_3$ ($h = 1.34\sigma$), assuming four different temperature values.  At low temperatures ({\it i.e.}, for $T \le T_{\rm s}$), the positions of the first three peaks of the in-plane correlation functions agree with high accuracy with the nearest neighbor distances in a planar, triangular lattice: for a given surface density,  $(H \rho_H) / 2$, the first three nearest neighbor distances are given by $d_1= a$, $d_2 = \sqrt{3} a$, and $d_3 = 2 a$, with the unit length of the triangular lattice, $a$, given by 

\begin{equation}
\label{eq:2Dtri}
a = \frac{2}{\sqrt{\sqrt{3}(H\rho_H)}} . 
\end{equation}

Taking the data accumulated in the Table, we obtain for the intermediate phase I$_2$ the value $a/\sigma=1.10735$; thus $d_1/\sigma = 1.10735$, $d_2/\sigma = 1.91799$, and $d_3/\sigma = 2.2147$. From the the in-plane correlation function $g_1(s)$ (shown in panel (a) of Figure \ref{fig:correlation_functions}), we observe for the positions of the first three peaks (and considering only the lower temperatures, $T= 0.05$ and $T = 0.15$) $s_1/\sigma = 1.109$, $s_2/\sigma = 1.910$ and $s_3/\sigma = 2.214$, {\it i.e.}, values that correspond with high accuracy to the predicted data. For the in-plane correlation function $g_2(s)$ (shown in panel (b) of Figure \ref{fig:correlation_functions}) the peak around $s/\sigma \sim 1.1$ is missing at low temperatures; however, it emerges as the temperature is increased.

Performing our analysis in a similar manner for the intermediate structure I$_3$, we obtain via Equation \ref{eq:2Dtri} the value $a/\sigma = 1.08826$. In contrast to the intermediate structure I$_2$, the main peaks of the in-plane correlation functions of the intermediate structure I$_3$, $g_1(s)$ and $g_2(s)$ ({\it cf.} panels (c) and (d) in Figure \ref{fig:correlation_functions}), occur for low temperatures at different positions. $g_1(s)$ shows at low temperatures a complex sequence of peaks: the main peak is located at $s/\sigma = 1.099$ [{\it i.e.}, a 5\% larger distance than the peak in $g_2(s)$] while the peaks corresponding to the second and the third neighbor distances are split into two sub-peaks. The position of the main peak in $g_2(s)$ corresponds to the nearest neighbor distance in a triangular lattice. Due to the characteristic differences in the positions of the main peaks in $g_2(s)$, we can distinguish between the two intermediate structures at low temperatures.

\begin{figure}
\begin{center}
\includegraphics[width=7cm]{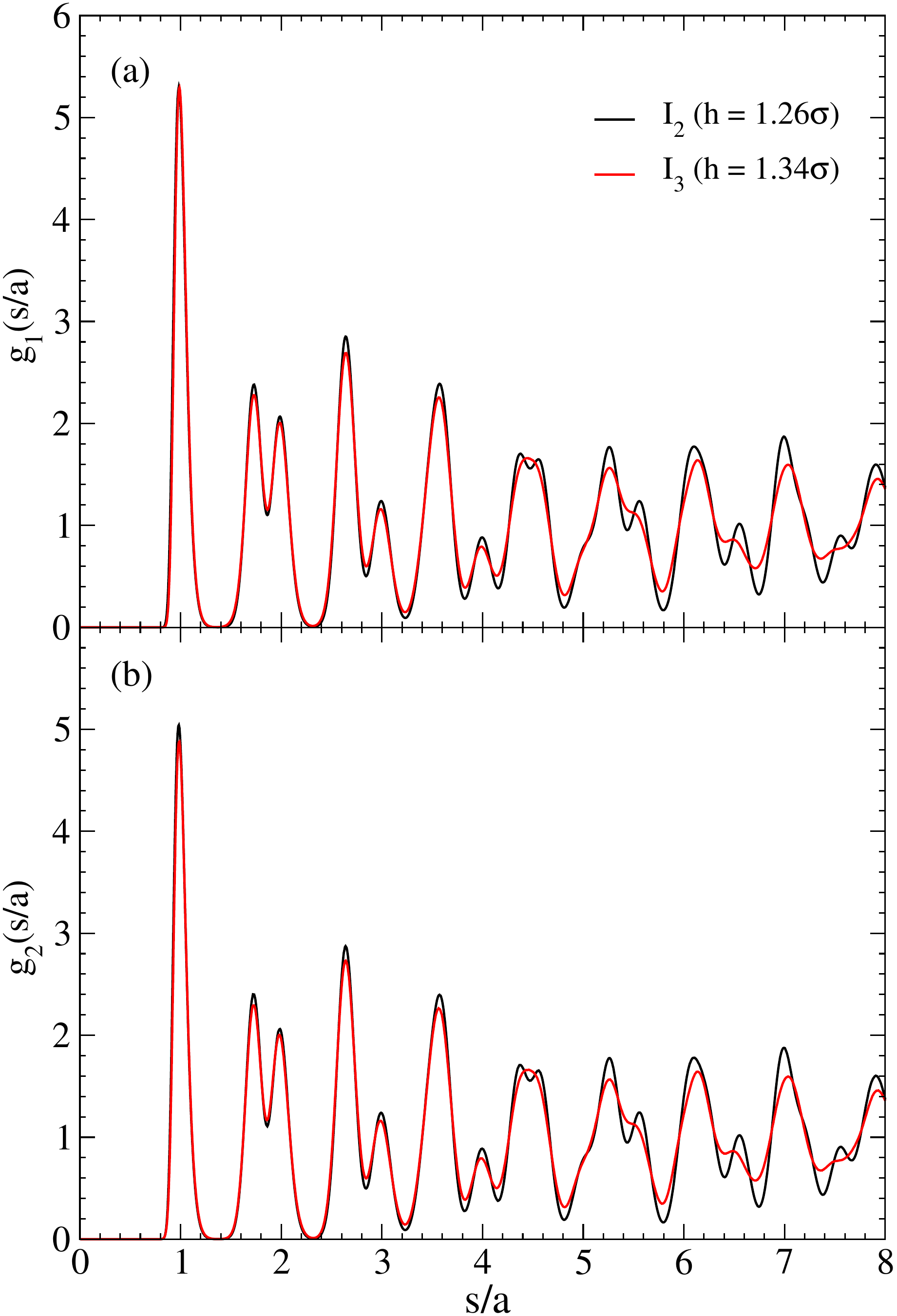}
\end{center}
\caption{\label{fig:g_scaled} Re-scaled in-plane correlation functions $g_1(s)$ and $g_2(s)$ for $n=12$, $\rho_H=1/1.2$, and $T = 1.0$ for the intermediate phases I$_2$ ($h=1.26 \sigma$) and I$_3$ ($h=1.34 \sigma$). Distances have been re-scaled with the unit length of the corresponding triangular lattice, $a$ [{\it cf.} Equation (\ref{eq:2Dtri})]. For $h = 1.26 \sigma$ we obtain $a/\sigma=1.10735$ while for $h = 1.34 \sigma$ we find $a/\sigma = 1.08826$. Panel (a) -- $g_1(s)$, panel (b) -- $g_2(s)$.}
\end{figure}

When the temperature is increased above $T_{\rm s}$, the system is in the 2$\triangle$ phase for both $h=1.26\sigma$ and $h=1.34\sigma$, as evidenced by the center-to-center correlation functions $g_1(s)$ and $g_2(s)$. Figure \ref{fig:g_scaled} shows these correlation functions for both slab widths, rescaled by the nearest neighbor distance in the respective triangular lattice; the resulting functions $g_1(s/a)$ and $g_2(s/a)$ are identical to the correlation function of a triangular lattice.

\section{Conclusions}
\label{sec:conclusions}

In this contribution we have identified the ordered ground state configurations ({\it i.e.}, at {\it vanishing} temperature) of soft particles confined between two impenetrable walls that are separated by a vertical distance $H$; furthermore we have investigated the mechanical stability at {\it finite} temperatures of some of the more complex particle arrangements.

The soft particles interact via an inverse power law potential (we assume exponents of 12, 21, and 30), guaranteeing thereby that their interactions are short-ranged. The ground state configurations have been identified via a highly reliable and efficient optimization tool that is based on ideas of evolutionary algorithms. Since we consider in these calculations up to 28 particle per unit cell, we are able to identify highly intricate ordered structures which occur predominantly as so-called intermediate phases: they typically emerge as the system transforms from the well-known layered, triangular to the layered, square structure (or vice versa). The latter particle arrangements are trivial and well-studied, being the consequence of the compatibility of the three dimensional fcc bulk lattice with a given $H$-value. In contrast, the intermediate structures are the outcome of the competition between two driving mechanisms: optimal packing of the particles (imposed by the incompatibility of the fcc bulk crystal with the respective $H$-value) versus energetical optimization. The result are highly complex layered structures, which are very often encountered in only very narrow $H$-intervals.  

In an effort to verify the mechanical stability of these intermediate phases at finite temperature, we have carried out extensive Monte Carlo simulations for a select set of these structures. Bridging a relatively broad temperature range with the help of multiple histogram reweighting methods we were able to identify in most cases a well-defined temperature up to which these structures are mechanically stable: our conclusions are based on a detailed analysis of the vertical density profiles and of the in-plane correlation functions obtained in the computer simulations. Taking benefit of the well-known scaling laws that are valid in inverse power law systems, we could show that suitable scaling laws can be applied both to the energetic as well as to the structural properties, even for the case that the particles are subject to confinement.

\section{Acknowledgements}
The authors acknowledge financial support from the projects PHC-Amadeus-2012/13 (project number 26996UC) and Projekt Amad\'ee (project number FR 10/2012), from the Austrian Research Foundation FWF (project number P23910-N16), and the SFB ViCoM (FWF-Spezialforschungsbereich F41). Part of the computational work has been performed within the HPC-EUROPA2 project (project number 228398) with the support of the European Commission - Capacities Area - Research Infrastructures. Additional computational resources were provided by the Direction Informatique of the Universit\'e de Paris-Sud. Special thanks is due to Erdal O\u{g}uz and Hartmut L\"{o}wen (both D\"usseldorf) for fruitful discussions and for sharing their data on hard sphere systems\cite{Oguz:09}.

\newpage

\begin{table*}
\caption{\label{table1} Most relevant input and output data of the MC simulations carried out for the four intermediate phases investigated ({\it i.e.}, I$_2$, I$_3$, I$_5$, and I$_6$). All simulations have been performed for $n=12$ and $\rho_H = 1/1.2$. $N_{\rm cell}$ is the number of primitive cells (whose structure was obtained with the optimization scheme) that is used to construct the initial configurations of the simulations; in parentheses we indicate the number of particles per primitive cell. $h$ is the effective slab width, $N$ is the total number of particles in the simulation box, $M$ is the number of slices used for the computation of the in-plane correlation functions [see Equation (\ref{eq:gm3})]. Further, the number of MC cycles for equilibration, $t_{\rm eq}$, and for accumulating the averages, $t_{\rm av}$, are listed. $T_{\rm s}$ indicates the temperature below which the corresponding phases are found to be mechanically stable (see text). $E_0/N\epsilon$ is the energy of the respective global ground state configurations as identified in our EA-based optimization scheme.}
\begin{ruledtabular}
\begin{tabular}{ccccccccc}
&&&&&&&&\\
phase & $N_{\rm cell}$ & $h/\sigma$ &  $N$ & $M$ & $t_{\rm eq}$ & $t_{\rm av}$ & $T_{\rm s}$ & $E_0/N\epsilon$ \\
\hline
I$_2$ & 252 (8)  & 1.26 & 2016 & 4 & $10^6$ & $2.8\times 10^6$ & 0.15 & 3.4835298 \\
I$_3$ & 144 (14) & 1.34 & 2016 & 4 & $8\times 10^5$ & $1.6\times 10^6$ & 0.25 & 3.8271842 \\
I$_5$ & 160 (18) & 2.25 & 2880 & 5 & $1.2\times 10^6$ & $1.6\times 10^6$ & $<$ 0.04 & 3.1008837 \\
I$_6$ & 100 (24) & 2.28 & 2400 & 5 & $1.2\times 10^6$ & $2\times 10^6$ & 0.04-0.05 & 3.2153247 \\
\end{tabular}
\end{ruledtabular}
\end{table*}

\newpage

\end{document}